\documentclass[onecolumn,showpacs,showkeys,preprintnumbers,amsmath,amssymb,a4paper,10pt]{revtex4}
\usepackage{amsthm}
\usepackage{mathrsfs}
\usepackage{graphicx}% Include figure files
\usepackage{dcolumn}% Align table columns on decimal point
\usepackage{bm}% bold math
\usepackage{subfigure}
\begin{document}

\title{Controllable entanglement preparations between atoms in spatially-separated cavities via quantum Zeno dynamics}

\author{Wen-An Li\footnote{E-mail: liwenan@126.com, liwa@mail.sysu.edu.cn}}
\affiliation{State Key Laboratory of Optoelectronic Materials
and Technologies, School of Physics and Engineering, Sun Yat-Sen
University Guangzhou 510275, China}
\author{L. F. Wei\footnote{E-mail: weilianfu@gmail.com, weilianf@mail.sysu.edu.cn}}
\affiliation{State Key Laboratory of Optoelectronic Materials
and Technologies, School of Physics and Engineering, Sun Yat-Sen
University Guangzhou 510275, China}
\affiliation{Quantum Optoelectronics Laboratory, School of Physics
and Technology, Southwest Jiaotong University, Chengdu 610031,
China}

\begin{abstract}
By using quantum Zeno dynamics, we propose a controllable approach to deterministically generate tripartite GHZ states
for three atoms trapped in spatially separated cavities.
The nearest-neighbored cavities are connected via optical fibers and the atoms
trapped in two ends are tunably driven.
The generation of the GHZ state can be implemented by only one step manipulation,
and the EPR entanglement between the atoms in two ends can be further realized deterministically by
Von Neumann measurement on the middle atom.
Note that the duration of the quantum Zeno dynamics is controllable by switching on/off the applied external classical drivings and the desirable tripartite GHZ state will no longer evolve once it is generated.
The robustness of the proposal is numerically demonstrated by considering various decoherence factors, including atomic spontaneous emissions, cavity decays and fiber photon leakages, etc. Our proposal can be directly generalized to generate multipartite entanglement by still driving the atoms in two ends.
\end{abstract}



\maketitle
\section{introduction}
Quantum entanglement, as a fundamental aspect in quantum mechanics, has occupied a central place in modern research because of its promise of enormous utility in quantum computing \cite{1,2,3}, cryptography \cite{2,4}, etc. Generally speaking, the more particles that can be entangled, the more clearly nonclassical effects are exhibited and the more useful the states are for quantum applications. Typically, tripartite GHZ state, first proposed by Greenberger, Horne, and Zeilinger, provides a possibility to test quantum mechanics against local hidden theory without inequality \cite{ghz} and has practical applications in e.g., quantum secrete sharing \cite{hbb}.

Recently, many theoretical schemes~\cite{ghz_t1,ghz_t2,ghz_t3,ghz_t4} have been proposed to generate multipartite GHZ states,
 and a series of experimental preparations~\cite{ghz_e1,ghz_e2,ghz_e3} have already been realized.
The physical systems utilized to these generations include superconducting circuits~\cite{ghz_t4,ghz_e3}, trapped ions~\cite{ghz_e2}, and cavity QED systems, etc..
It is well-known that cavity QED, where atoms interact with quantized electromagnetic fields inside a cavity, is a useful platform to demonstrate fundamental quantum mechanics laws and for  the implementation of quantum information processing~\cite{qed}.
Specifically, numerous proposals with cavity QED have been made for entangling atoms either inside a cavity \cite{cz,hl} or trapped individually in different cavities~\cite{czkm,ecz,bkpv,lssh,pk}. For example, Pellizzari \cite{p} proposed an approach to realize the reliable transfer of quantum information between two atoms in distant cavities {\it connected by an optical fiber}. Based on this proposal, various schemes \cite{19,20,21,22,23,24,25,26,27,zl} have been proposed to realize the quantum manipulations of the atoms trapped in different cavities connected via optical fibers. However, these models either require strong cavity-fiber coupling~\cite{24} (which is not easy to achieve in the usual experiment), or need many laser beams to implement the desirable manipulations~\cite{25} (this increases the experimental complication), or is sensitive to the decays of cavity fields and fiber modes~\cite{zl} (which limits its scalability).

Here, by using quantum Zeno dynamics we propose an alternative approach to entangle the atoms trapped in the distant
 cavities connected by optical fibers. It is well-known that quantum Zeno effect is an interesting phenomenon
  in quantum mechanics. Due to this effect the quantum system remains in its initial state via frequently measurements.
   Facchi et al \cite{zeno1,zeno2,zeno3} showed that this effect does not necessarily freeze the dynamics.
    Instead, by frequently projecting onto a multidimensional subspace, the system could evolve away from its
     initial state, although it remains in the so-called ``Zeno subspace''~\cite{zeno1,zeno4}.
     Moreover, without making use of projection operators and nonunitary dynamics, the quantum Zeno effect can
     also be expressed in terms of a continuous coupling between the system and detector~\cite{zeno3}.
Generally, the system and its continuously coupling detector can be governed by the total Hamiltonian $H_K=H+KH_a$, with $H$ is for the quantum system investigated and the $H_a$ describing the additional interaction with the detector, and $K$ the coupling constant. In the limit $K\rightarrow\infty$, the subsystem of interest is dominated by the evolution operator $U(t)=\lim_{K\rightarrow\infty}\exp(iKH_at)U_K(t)$ which takes the form $U(t)=\exp(-it\sum_n P_nHP_n)$~\cite{zeno3}. Here, $P_n$ is the eigenprojection of $H_a=\sum_n \lambda_n P_n$ corresponding to the eigenvalue $\lambda_n$. As a consequence, the system-detector can be described by the evolution operator $U_K(t)\sim\exp(-iKH_at)U(t)=\exp[-i\sum_n(K\lambda_n P_n+P_nHP_n)t]$. This result is of great importance in view of practical applications of the quantum Zeno dynamics, such as to prepare various quantum states~\cite{state1,state2,me} and to implement the quantum gates~\cite{gate1,gate2,gate3,gate4}.

Compared with previous protocols~\cite{24,25,zl} for generating GHZ states by selective absorption and emission of photons, adiabatic passages, and dispersive interactions between the atoms and cavities, etc., our approach possesses the following advantages: (i) the expected entanglement can be established by only one step operation, (ii) it is robust with respect to parameter imprecision and atomic and fibers' dissipations, (iii) under the Zeno condition, the cavity fields are not excited really and thus is insensitive to the decays of the cavities, (iv) the generalization to N-atom entanglement is direct, and no matter how many atoms are involved, two laser beams are enough to implement the generations.

The paper is organized as follows. In section II, we present our generic approach by using quantum Zeno dynamics to implement the GHZ entanglement of the atoms trapped in three fiber-connected cavities. The validity of the used quantum Zeno dynamics is analyzed by numerical method in detail. The direct generalization for entangling the $N$ atoms trapped individually in $N$ cavity is provided in Sec. III. Finally, in Sec. IV we discuss the feasibility of our proposal and give our conclusions.

\section{Generation of GHZ state of atoms trapped in different cavities by quantum Zeno dynamics}
We consider the physical configuration shown in the Fig.~1, three $\Lambda$-type atoms are trapped in three distant optical cavities coupled by two short optical fibers. Each atom has one excited state $|e\rangle$ and two dipole-transition forbidden ground states $|0\rangle$, $|1\rangle$. The first and third atomic transitions $|1\rangle\leftrightarrow|e\rangle$ are driven resonantly by classical lasers with the couplings coefficient $\Omega_1$ and $\Omega_3$, respectively. The other atomic transition is resonantly coupled to the corresponding cavity mode with coupling constant $g_{i,r(l)}$ ($i=1,2,3$). The subscript $r(l)$ denotes the right (left) circularly polarization. In the short fiber limit, $(2L\bar{v})/(2\pi c)\ll 1$, where $L$ is the length of the fiber and $\bar{v}$ is the decay rate of the cavity fields into a continuum of fiber modes~\cite{19}. In the interaction picture, the Hamiltonian of the whole system can be written as
\begin{figure}[t]
\begin{center}
\includegraphics[width=0.6\textwidth]{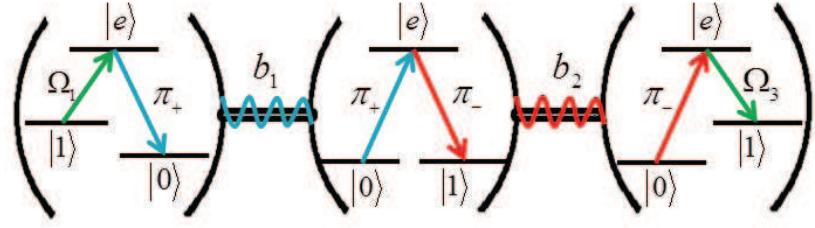}
\end{center}
\caption{Zeno manipulations of atoms in spatially-separated cavities
connected via optical fibers. Here, $\Omega_1$ ($\Omega_3$) is the classical field coupled to the first (third) atom, and $b_1$ and $b_2$ are the bosonic operators in fibers and couple to the corresponding cavity modes.}
\end{figure}

\begin{equation}\label{11}
H_{\mathrm{total}}=H_l+H_{a-c-f},
\end{equation}
\begin{equation}
H_l=\Omega_1 |e\rangle_1\langle 1|+\Omega_3 |e\rangle_3\langle 1|+\mathrm{h.c.},
\end{equation}
\begin{eqnarray}
\nonumber H_{a-c-f}&=&g_{1,r} a_{1,r}|e\rangle_1\langle 0|+g_{2,r} a_{2,r}|e\rangle_2\langle 0|+g_{2,l} a_{2,l}|e\rangle_2\langle 1|+g_{3,l} a_{3,l}|e\rangle_3\langle 0|\\
&+&v_1b_1^\dag(a_{1,r}+a_{2,r})+v_2b_2^\dag(a_{2,l}+a_{3,l})+\mathrm{h.c.},
\end{eqnarray}
where $a$ and $b$ are the annihilation operators associated with the modes of cavity and fiber respectively and $v_i$($i=1,2$) is the corresponding cavity-fiber coupling constant. We assume that $g_{1,r}=g_{2,r(l)}=g_{3,l}=g$ and $v_1=v_2=v$ for convenience.
If the initial state of the whole system is $|1,0,0\rangle_a|0\rangle_{c_1}|0\rangle_{f_1}|0,0\rangle_{c_2}|0\rangle_{f_2}|0\rangle_{c_3}$, then the evolution of system will be restricted in the subspace spanned by
\begin{eqnarray}\label{33}
\nonumber  |\phi_1\rangle &=& |1,0,0\rangle_a|0\rangle_{c_1}|0\rangle_{f_1}|0,0\rangle_{c_2}|0\rangle_{f_2}|0\rangle_{c_3}, \quad
|\phi_2\rangle = |e,0,0\rangle_a|0\rangle_{c_1}|0\rangle_{f_1}|0,0\rangle_{c_2}|0\rangle_{f_2}|0\rangle_{c_3}, \\
\nonumber  |\phi_3\rangle &=& |0,0,0\rangle_a|1\rangle_{c_1}|0\rangle_{f_1}|0,0\rangle_{c_2}|0\rangle_{f_2}|0\rangle_{c_3}, \quad
|\phi_4\rangle = |0,0,0\rangle_a|0\rangle_{c_1}|1\rangle_{f_1}|0,0\rangle_{c_2}|0\rangle_{f_2}|0\rangle_{c_3}, \\
\nonumber  |\phi_5\rangle &=& |0,0,0\rangle_a|0\rangle_{c_1}|0\rangle_{f_1}|1,0\rangle_{c_2}|0\rangle_{f_2}|0\rangle_{c_3}, \quad
  |\phi_6\rangle = |0,e,0\rangle_a|0\rangle_{c_1}|0\rangle_{f_1}|0,0\rangle_{c_2}|0\rangle_{f_2}|0\rangle_{c_3}, \\
\nonumber  |\phi_7\rangle &=& |0,1,0\rangle_a|0\rangle_{c_1}|0\rangle_{f_1}|0,1\rangle_{c_2}|0\rangle_{f_2}|0\rangle_{c_3}, \quad
|\phi_8\rangle = |0,1,0\rangle_a|0\rangle_{c_1}|0\rangle_{f_1}|0,0\rangle_{c_2}|1\rangle_{f_2}|0\rangle_{c_3}, \\
\nonumber  |\phi_9\rangle &=& |0,1,0\rangle_a|0\rangle_{c_1}|0\rangle_{f_1}|0,0\rangle_{c_2}|0\rangle_{f_2}|1\rangle_{c_3}, \quad
|\phi_{10}\rangle = |0,1,e\rangle_a|0\rangle_{c_1}|0\rangle_{f_1}|0,0\rangle_{c_2}|0\rangle_{f_2}|0\rangle_{c_3}, \\
|\phi_{11}\rangle &=& |0,1,1\rangle_a|0\rangle_{c_1}|0\rangle_{f_1}|0,0\rangle_{c_2}|0\rangle_{f_2}|0\rangle_{c_3},
\end{eqnarray}
where $|i,j,k\rangle_a$\,($i,j,k=0,1,e$) denotes the state of atoms in each cavity, $n$ in $|n\rangle_s$\,($s=c_1, f_1, c_2, f_2, c_3$) denotes the photon number in cavities or fibers.

Under the Zeno condition $g,v\gg \Omega_1,\Omega_3$, the above
Hilbert subspace is split into nine invariant Zeno subspaces
\cite{zeno3}
\begin{eqnarray}
\nonumber  \Gamma_{P_1} &=& \{|\phi_1\rangle, |\psi_1\rangle, |\phi_{11}\rangle\}, \quad \Gamma_{P_2}=\{|\psi_2\rangle\}, \\
  \Gamma_{P_3} &=& \{|\psi_3\rangle\}, \quad \Gamma_{P_4} = \{|\psi_4\rangle\}, \quad \Gamma_{P_5} = \{|\psi_5\rangle\}, \quad \Gamma_{P_6} = \{|\psi_6\rangle\},\\
\nonumber  \Gamma_{P_7} &=& \{|\psi_7\rangle\}, \quad \Gamma_{P_8} = \{|\psi_8\rangle\}, \quad \Gamma_{P_9} = \{|\psi_9\rangle\}
\end{eqnarray}
corresponding to the projections
\begin{equation}
P_i^\alpha=|\alpha\rangle \langle \alpha|, \quad (|\alpha\rangle\in\Gamma_{P_i})
\end{equation}
with eigenvalues $\lambda_1=0$, $\lambda_2=-\sqrt{(g^2+2v^2-A)/2}$, $\lambda_3=\sqrt{(g^2+2v^2-A)/2}$, $\lambda_4=-\sqrt{(3g^2+2v^2-A)/2}$, $\lambda_5=\sqrt{(3g^2+2v^2-A)/2}$, $\lambda_6=-\sqrt{(3g^2+2v^2+A)/2}$, $\lambda_7=\sqrt{(3g^2+2v^2+A)/2}$, $\lambda_8=-\sqrt{(g^2+2v^2+A)/2}$, $\lambda_9=\sqrt{(g^2+2v^2+A)/2}$, where $A=\sqrt{g^4+4v^4}$.
Here,
\begin{eqnarray}
\nonumber  |\psi_1\rangle &=& N_1\left(|\phi_2\rangle-\frac{g}{v}|\phi_4\rangle+|\phi_6\rangle-\frac{g}{v}|\phi_8\rangle+|\phi_{10}\rangle\right) \\
\nonumber  |\psi_2\rangle &=& N_2\left(-|\phi_2\rangle+\epsilon_1|\phi_3\rangle-\eta_1|\phi_4\rangle-\chi_1|\phi_5\rangle+\chi_1|\phi_7\rangle+\eta_1|\phi_8\rangle-\epsilon_1|\phi_9\rangle+|\phi_{10}\rangle\right) \\
\nonumber  |\psi_3\rangle &=& N_3\left(-|\phi_2\rangle-\epsilon_1|\phi_3\rangle-\eta_1|\phi_4\rangle+\chi_1|\phi_5\rangle-\chi_1|\phi_7\rangle+\eta_1|\phi_8\rangle+\epsilon_1|\phi_9\rangle+|\phi_{10}\rangle\right) \\
\nonumber  |\psi_4\rangle &=& N_4\left(|\phi_2\rangle-\mu_1|\phi_3\rangle-\zeta_1|\phi_4\rangle+\delta_1|\phi_5\rangle-\theta_1|\phi_6\rangle+\delta_1|\phi_7\rangle-\zeta_1|\phi_8\rangle-\mu_1|\phi_9\rangle+|\phi_{10}\rangle\right) \\
  |\psi_5\rangle &=& N_5\left(|\phi_2\rangle+\mu_1|\phi_3\rangle-\zeta_1|\phi_4\rangle-\delta_1|\phi_5\rangle-\theta_1|\phi_6\rangle-\delta_1|\phi_7\rangle-\zeta_1|\phi_8\rangle+\mu_1|\phi_9\rangle+|\phi_{10}\rangle\right) \\
\nonumber  |\psi_6\rangle &=& N_6\left(-|\phi_2\rangle+\epsilon_2|\phi_3\rangle-\eta_2|\phi_4\rangle+\chi_2|\phi_5\rangle-\chi_2|\phi_7\rangle+\eta_2|\phi_8\rangle-\epsilon_2|\phi_9\rangle+|\phi_{10}\rangle\right) \\
\nonumber  |\psi_7\rangle &=& N_7\left(-|\phi_2\rangle-\epsilon_2|\phi_3\rangle-\eta_2|\phi_4\rangle-\chi_2|\phi_5\rangle+\chi_2|\phi_7\rangle+\eta_2|\phi_8\rangle+\epsilon_2|\phi_9\rangle+|\phi_{10}\rangle\right) \\
\nonumber  |\psi_8\rangle &=& N_8\left(|\phi_2\rangle-\mu_2|\phi_3\rangle+\zeta_2|\phi_4\rangle-\delta_2|\phi_5\rangle+\theta_2|\phi_6\rangle-\delta_2|\phi_7\rangle+\zeta_2|\phi_8\rangle-\mu_2|\phi_9\rangle+|\phi_{10}\rangle\right) \\
\nonumber  |\psi_9\rangle &=& N_9\left(|\phi_2\rangle+\mu_2|\phi_3\rangle+\zeta_2|\phi_4\rangle+\delta_2|\phi_5\rangle+\theta_2|\phi_6\rangle+\delta_2|\phi_7\rangle+\zeta_2|\phi_8\rangle+\mu_2|\phi_9\rangle+|\phi_{10}\rangle\right)
\end{eqnarray}
with
\begin{eqnarray}
\nonumber  \epsilon_1&=&\frac{\sqrt{g^2+2v^2-A}}{\sqrt{2}g}, \quad \eta_1=\frac{-g^2+2v^2-A}{2gv}, \quad \chi_1=\frac{\sqrt{g^2+2v^2-A}(g^2+A)}{2\sqrt{2}gv^2},\\
\nonumber  \mu_1&=&\frac{\sqrt{3g^2+2v^2-A}}{\sqrt{2}g}, \quad \zeta_1=\frac{-g^2-2v^2+A}{2gv}, \quad \delta_1=\frac{\sqrt{3g^2+2v^2-A}(-g^2+A)}{2\sqrt{2}gv^2},
  \quad \theta_1=\frac{-g^2+A}{v^2},\\ \nonumber
  \epsilon_2&=&\frac{\sqrt{g^2+2v^2+A}}{\sqrt{2}g}, \quad \eta_2=\frac{-g^2+2v^2+A}{2gv}, \quad \chi_2=\frac{\sqrt{g^2+2v^2+A}(-g^2+A)}{2\sqrt{2}gv^2},\\
  \mu_2&=&\frac{\sqrt{3g^2+2v^2+A}}{\sqrt{2}g}, \quad \zeta_2=\frac{g^2+2v^2+A}{2gv}, \quad \delta_2=\frac{\sqrt{3g^2+2v^2+A}(g^2+A)}{2\sqrt{2}gv^2},
  \quad \theta_2=\frac{g^2+A}{v^2},
\end{eqnarray}
and $N_i$\,($i=1,2,3,...,9$) being the normalization factor for the
eigenstate $|\psi_i\rangle$.
Under above condition, the system can be effectively described by
the following Hamiltonian
\begin{eqnarray}
\nonumber  H_{\mathrm{total}} &\simeq& \sum_{i,\alpha,\beta}\lambda_i P_i^{\alpha}+P_i^{\alpha}H_lP_i^{\beta} \\
  &=&\sum_{i=2}^9 \lambda_i|\psi_i\rangle \langle\psi_i|+N_1(\Omega_1|\phi_1\rangle \langle\psi_1|+\Omega_3|\phi_{11}\rangle \langle\psi_1|+\mathrm{h.c.})
\end{eqnarray}
It reduces to
\begin{equation}
H_{\mathrm{eff}}=N_1(\Omega_1|\phi_1\rangle \langle\psi_1|
+\Omega_3|\phi_{11}\rangle \langle\psi_1|+\mathrm{h.c.}),
\end{equation}
if the initial state is $|\phi_1\rangle$.
The effective Hamiltonian $H_{\mathrm{eff}}$ implies that the
evolution of system is restricted in the subspace, wherein the
cavity modes are kept in the vacuum. Consequently, after the
evolution time $t$ the state of the system becomes
\begin{eqnarray}\label{22}
\nonumber|\Psi(t)\rangle&=&\frac{\cos(N_1t\sqrt{\Omega_1^2+\Omega_3^2})\Omega_1^2+\Omega_3^2}{\Omega_1^2+\Omega_3^2}|\phi_1\rangle - \frac{i\Omega_1\sqrt{\Omega_1^2+\Omega_3^2}\sin(N_1t\sqrt{\Omega_1^2+\Omega_3^2})}{\Omega_1^2+\Omega_3^2}|\psi_1\rangle\\
 &+& \frac{[\cos(N_1t\sqrt{\Omega_1^2+\Omega_3^2})-1]\Omega_1\Omega_3}{\Omega_1^2+\Omega_3^2}|\phi_{11}\rangle.
\end{eqnarray}
Obviously, if $\Omega_1=(\sqrt{2}+1)\Omega_3$ and the evolution time
is set as $t=\tau=\pi/N_1\sqrt{\Omega_1^2+\Omega_3^2}$, then the
triatomic GHZ state $|\Psi\rangle_a$ can be generated, i.e.,
\begin{equation}
|\Psi(\tau)\rangle=-\frac{1}{\sqrt{2}}(|\phi_1\rangle+|\phi_{11}\rangle)
=|\Psi\rangle_a\otimes|0\rangle_{c_1}|0\rangle_{f_1}|0,0\rangle_{c_2}
|0\rangle_{f_2}|0\rangle_{c_3},\,\,
|\Psi\rangle_a=-\frac{1}{\sqrt{2}}(|1,0,0\rangle_a+|0,1,1\rangle_a).
\end{equation}
It is emphasized that the duration $\tau$ depends directly on the Rabi frequencies $\Omega_1$ and $\Omega_3$
applied simultaneously. Thus, the above quantum Zeno dynamics stops once the classical fields $\Omega_1$
and $\Omega_3$ are switched off simultaneously. Fortunately, one can easily check that
$H_{a-c-f}|\Psi(\tau)\rangle=0$, which means that generated GHZ state does not evolve
when the controllable quantum Zeno dynamics vanishes.
\begin{figure}[t]
\begin{center}
\includegraphics[width=0.5\textwidth]{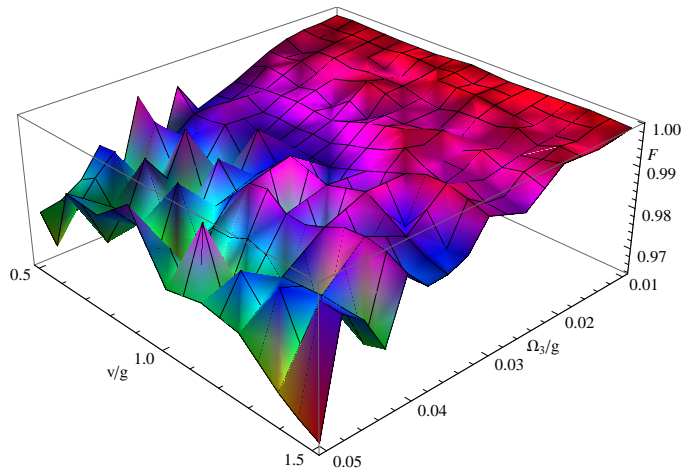}
\end{center}
\caption{The influence of the ratios: $\Omega_3/g$ and $v/g$, on the fidelity of the prepared GHZ state.}
\end{figure}
Furthermore, if we perform a single-qubit rotation $R_x(\pi/4)(=\exp(i\sigma_x\cdot \pi/4))$ on the middle atom, then the tripartite GHZ state reduces to the state: $[(|1,0\rangle_{1,3}+i|0,1\rangle_{1,3})|0\rangle_2+(i|1,0\rangle_{1,3}+|0,1\rangle_{1,3})|1\rangle_2]/2$. Consequently, the EPR entanglement between two distant atoms can be deterministically obtained by projective measurement on the middle atom (no matter it is found at which state)~\cite{wei-prb-2005}, without taking any operation on the first and third atom directly.

It is clear that the validity of our scheme mainly relies on if the
so-called Zeno condition $g,v\gg \Omega_1,\Omega_3$ is satisfied
robustly. Now, we discuss how the ratio
$\Omega_3/g$ influences the fidelity $F=|\langle
\Psi(\tau)|\varphi(\tau)\rangle|^2$, with $|\varphi(\tau)\rangle$
being the relevant final state evolved by the original
$H_{\mathrm{total}}$ defined in Eq.(\ref{11}).
On the other hand, the ratio $v/g$ is another important factor
affecting the fidelity~\cite{19,20,21,24}.
Indeed, from Fig.~2 we see that the smaller the ratio $\Omega_3/g$
corresponds to the higher fidelity. We also note that, even at the
relatively-low ratio $v/g=0.5$, the fidelity is still high, i.e.,
above 97\%. This is very important for the practical application of our scheme,
as the large cavity-fiber coupling is not easy to be satisfied in the
realistic experiments. Furthermore, in order to obtain high fidelity in moderate
time, we can choose typically the ratios: $\Omega_3/g=0.04$ and
$v/g=1$.

Our proposal is also robust for the imprecision of experimental
parameters. Indeed, our previous discussions are based on certain
ideal assumptions, e.g., $g_{i,r(l)}\equiv g$, $v_i\equiv v$.
Practically, the coupling strengths $g_{i,r(l)}$ depend on the
positions of the atoms in the cavities, and thus certain deviations
are unavoidable. To numerically consider these deviations, we
typically set $g_{2,r(l)}=g$, $g_{1,r}=g+\delta g_1$ and
$g_{3,l}=g+\delta g_3$ for simulations. In Fig.~3(a), the fidelity
of the prepared state versus the variation $\delta g_1$ and $\delta
g_3$ is plotted.
\begin{figure}[b]
\begin{center}
\subfigure[]{\includegraphics[width=0.4\textwidth]{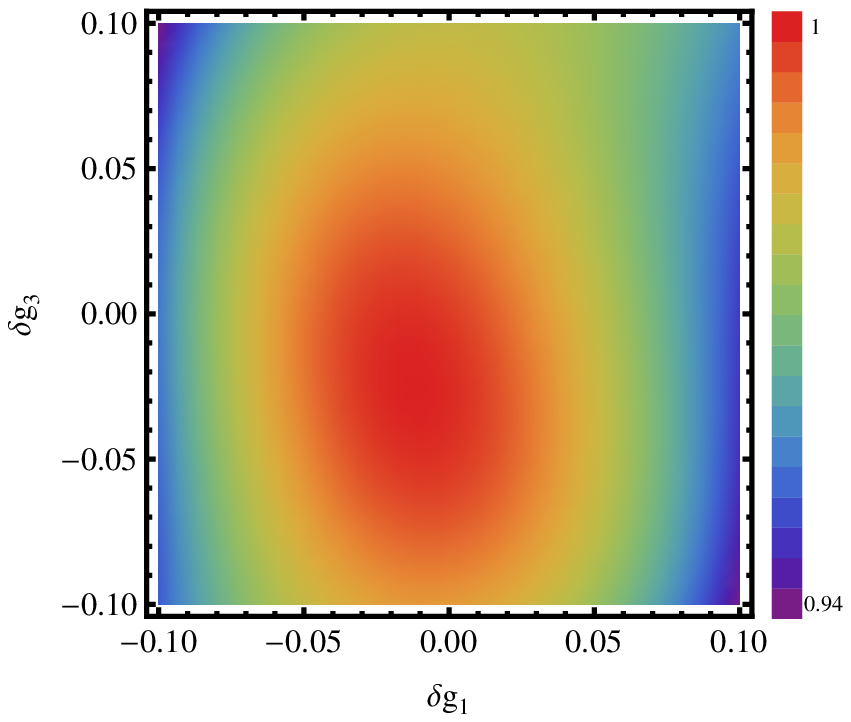}}
\subfigure[]{\includegraphics[width=0.4\textwidth]{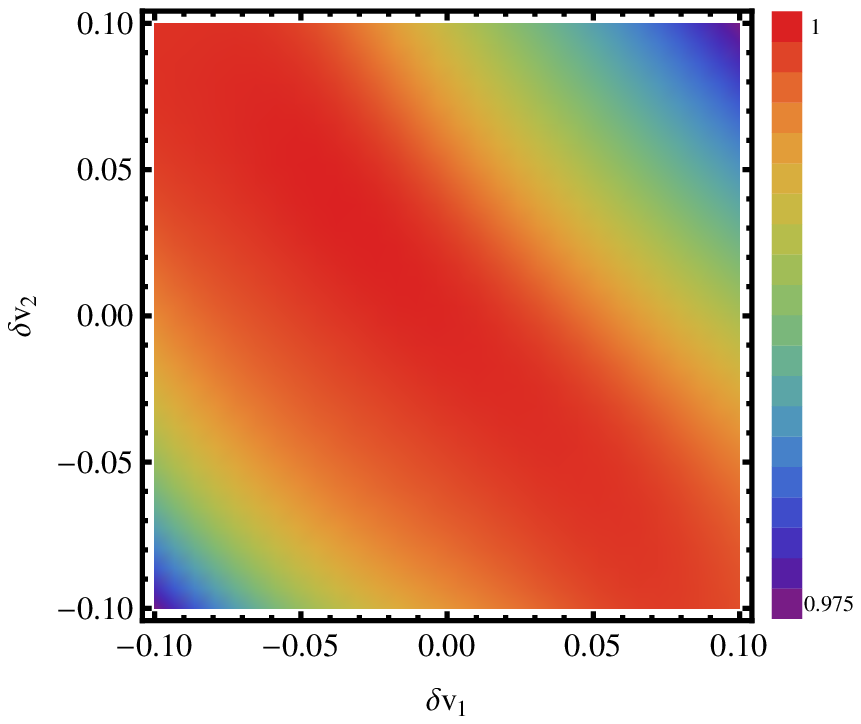}}
\end{center}
\caption{The fidelity of the three-atom GHZ state versus various parameter errors: (a) $\delta g_1$ and $\delta g_3$; (b) $\delta v_1$ and $\delta v_2$.}
\end{figure}
It is seen that, even under a deviation $|\delta
g_{1,3}|=10\% g$, the fidelity is still sufficiently high, e.g.,
larger than 94\%. Similarly, we plot the fidelity versus the
deviation of the cavity-fiber coupling ($v$) in Fig.~3(b), and find
also that the great robustness against the errors in the selected
parameters.

We now numerically verify that the above effective dynamics
restricted in the subspace without exciting the cavity modes is also
robust. In fact, considering all possible states of the system, the
state at the time $t$ reads $|\varphi_{total}\rangle=\sum_i
c_i(t)|\phi_i\rangle$ within the subspace spanned by the basic state
vectors in Eq.~(\ref{33}). The occupation probability for each state
vector $|\phi_i\rangle$ during the evolution is $P_i(t)=|c_i(t)|^2$,
and satisfies the condition $\sum_iP_i(t)=1$. By solving the
Schr\"{o}dinger equation, the variation of the occupation
probability $P_c(t)$ ($\equiv \sum_{i=3, 5, 7, 9}P_i(t)$) is
portrayed in Fig.~4 (red line).
\begin{figure}
\begin{center}
\includegraphics[width=0.5\textwidth]{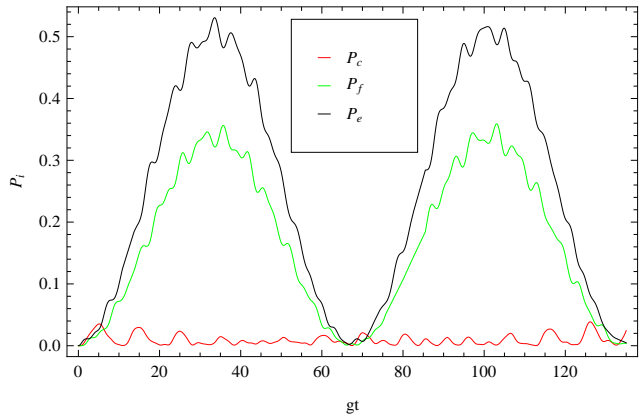}
\end{center}
\caption{The population probabilities of cavity photonic state $P_c$, fiber photonic state $P_f$ and the atomic excited state $P_e$ versus the dimensionless parameter $gt$ by exactly solving the evolution equation of the system without any approximation.}
\end{figure}
As shown in Fig.~4, the occupation probability of the cavity photonic
state is less than $0.04$. Therefore, our claim that the scheme is
immune to the cavity decay seems justifiable. Meanwhile, Fig.~4 also
describes the occupation probabilities $P_f(t)$($\equiv \sum_{i=4,
8}P_i(t)$, green line) and $P_e(t)$($\equiv \sum_{i=2, 6,
10}P_i(t)$, black line) for those states, wherein the photon is in the
fiber and one of the atoms in its excited state, respectively. Since $P_e>P_f$
in Fig.~4, our protocol is more sensitive to atomic spontaneous
emission than the fiber loss. To check this,
let us investigate the evolution of the system governed by the
following non-Hermitian Hamiltonian
\begin{equation}
H_{\mathrm{dec}}=H_{\mathrm{total}}-\frac{i\gamma}{2}\sum_{i=1}^3 |e\rangle_i\langle e|-\frac{i\kappa_c}{2}\left(\sum_{i=1,2}a^\dag_{i,r}a_{i,r}+\sum_{i=2,3}a^\dag_{i,l}a_{i,l}\right)-\frac{i\kappa_f}{2}\sum_{j=1,2}b_i^\dag b_i,
\end{equation}
where $\gamma$ is the spontaneous emission rate for atoms and $\kappa_{c(f)}$ denotes the decay rate of the cavity modes (fiber modes).
\begin{figure}[t]
\begin{center}
\includegraphics[width=0.5\textwidth]{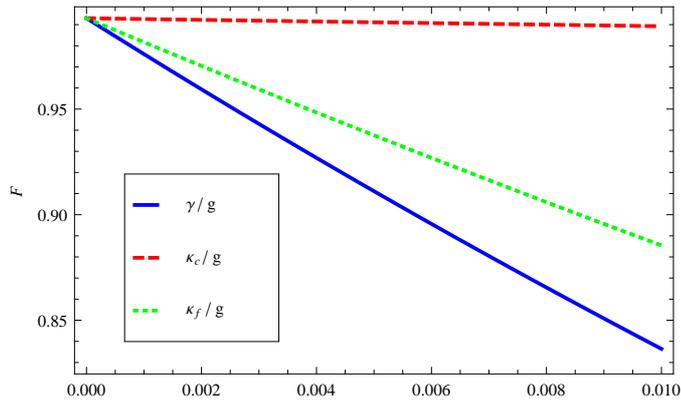}
\end{center}
\caption{The influences of atomic spontaneous emission $\gamma/g$, cavity field decay $\kappa_c/g$ and fiber photonic leakage $\kappa_f/g$ on the fidelity of the triatomic GHZ state for the typical ratios: $\Omega_3/g=0.04$ and $v/g=1$.}
\end{figure}
It is seen from Fig.~5 that the atomic spontaneous emission is the dominant factor
of degrading the fidelity of the generated GHZ state. This can also
be seen directly in the explicit form of the state $|\psi_1\rangle$,
where the population probability of the atomic excited state is larger than that of the photon in fiber.

\section{Generalization to N-atom entanglement}
We now generalize the present scheme to generate $N$-atom GHZ states. Let us consider the configuration shown in Fig.~6,
where $N$ atoms are individually trapped in $N$ cavities connected by $N-1$ short fibers.
The level configuration of the atoms between two ends are chosen the same as that of the middle
atom in the above 3-atom case.
The Hamiltonian of the present system reads
\begin{equation}\label{total-n}
H^\prime_{total}=H^\prime_l+H^\prime_{\mathrm{a-c-f}}
\end{equation}
where
\begin{equation}
H^\prime_l=\Omega_1|e\rangle_1\langle 1|+\Omega_N|e\rangle_N\langle 1|+\mathrm{h.c.}
\end{equation}
\begin{figure}[b]
\begin{center}
\includegraphics[width=0.6\textwidth]{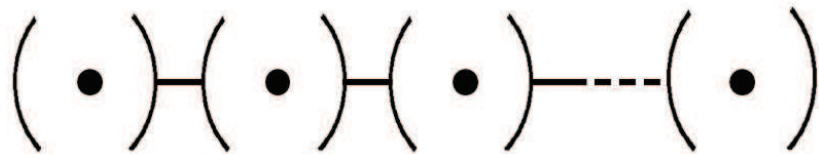}
\end{center}
\caption{$N$ atom trapped in distant cavities.}
\end{figure}
\begin{eqnarray}
\nonumber  H^\prime_{\mathrm{a-c-f}}&=& g_1a_{1,r}|e\rangle_1\langle 0|+\sum_{j=1}^k (g_{2j,r}a_{2j,r}|e\rangle_{2j}\langle 0|+g_{2j,l}a_{2j,l}|e\rangle_{2j}\langle 1|) \\ \nonumber
  &+&\sum_{j=1}^{k-1} (g_{2j+1,r}a_{2j+1,r}|e\rangle_{2j+1}\langle 1|+g_{2j+1,l}a_{2j+1,l}|e\rangle_{2j+1}\langle 0|)+g_Na_{N,l}|e\rangle_1\langle 0|\\
  &+&\sum_{j=1}^k[v_{2j-1}b_{2j-1}^+(a_{2j-1,r}+a_{2j,r})+v_{2j}b_{2j}^+(a_{2j,l}+a_{2j+1,l})]+\mathrm{h.c.}.
\end{eqnarray}
Without loss of generality, we consider the case where $N$ is odd number ($N=2k+1,\,k=1,2,3,...$), and
$g_{i,r(l)}=g$, $v_i=v$. If the initial state of the whole system is prepared at the state
$|1,0,0,...,0\rangle_a|0\rangle_{all}$, then the system will evolve within the subspace $\Gamma_{\mathrm{full}}$
spanned by the vectors: $\{|\phi^\prime_1\rangle,|\phi^\prime_2\rangle,|\phi^\prime_3\rangle,...,|\phi^\prime_{8k+3}\rangle\}$:
\begin{figure}
\begin{center}
\includegraphics[width=0.5\textwidth]{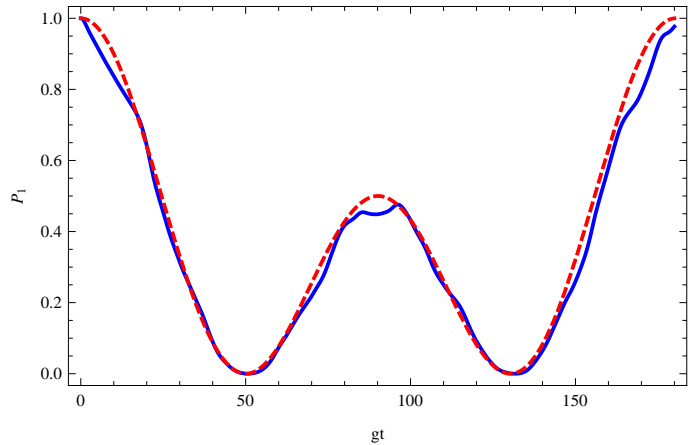}
\end{center}
\caption{(Color online) The occupation probabilities of the state $|\phi'_1\rangle$ versus the dimensionless parameter $gt$ under the total Hamiltonian (solid-line) and effective Hamiltonian (dotted-line).}
\end{figure}
\begin{eqnarray}\label{sp}
\nonumber|\phi^\prime_1\rangle &=& |1,0,0,...,0\rangle_a|0\rangle_{all},\quad
|\phi^\prime_2\rangle = |e,0,0,...,0\rangle_a|0\rangle_{all},\quad
|\phi^\prime_3\rangle = |0,0,0,...,0\rangle_a|1\rangle_{c_1},\\
\nonumber|\phi^\prime_4\rangle &=& |0,0,0,...,0\rangle_a|1\rangle_{f_1},\quad
|\phi^\prime_5\rangle = |0,0,0,...,0\rangle_a|1,0\rangle_{c_2},\quad
|\phi^\prime_6\rangle = |0,e,0,...,0\rangle_a|0\rangle_{all},\\
|\phi^\prime_7\rangle &=& |0,1,0,...,0\rangle_a|0,1\rangle_{c_2},\quad
|\phi^\prime_8\rangle = |0,1,0,...,0\rangle_a|1\rangle_{f_2},\quad
|\phi^\prime_9\rangle = |0,1,0,...,0\rangle_a|0,1\rangle_{c_3},\\
\nonumber|\phi^\prime_{10}\rangle &=& |0,1,e,...,0\rangle_a|0\rangle_{all},\quad
...\quad
|\phi^\prime_{8k+3}\rangle = |0,1,1,...,1\rangle_a|0\rangle_{all}
\end{eqnarray}
where $|0\rangle_{all}$ means that all boson modes are in the vacuum state, $n$\,($n_1$,\,$n_2$) in $|n\rangle_s$ or $|n_1,n_2\rangle_s$\,($s=c_i, f_i$) denotes the photon number in the corresponding resonator while other cavities or fibers are in the vacuum state. Similar to the above procedure, we get the effective Hamiltonian
\begin{equation}\label{eff-n}
H^\prime_{\mathrm{eff}}=N^\prime_1(\Omega_1|\phi^\prime_1\rangle \langle\psi^\prime_1|+\Omega_N|\phi^\prime_{8k+3}\rangle \langle\psi^\prime_1|+\mathrm{h.c.}),
\end{equation}
where
\begin{equation}
|\psi^\prime_1\rangle=N^\prime_1\left(\sum_{i=1}^N|\phi^\prime_{4i-2}\rangle
-\sum_{i=1}^{N-1}\frac{g}{v}|\phi^\prime_{4i}\rangle\right).
\end{equation}
Set $\Omega_1=(\sqrt{2}+1)\Omega_N$ and the interaction time $\tau^\prime=\pi/N_1^\prime\sqrt{\Omega_1^2+\Omega_N^2}$, the desirable $N$-atomic GHZ state
\begin{equation}
|\Psi(\tau^\prime)\rangle=-\frac{1}{\sqrt{2}}(|\phi^\prime_1\rangle+|\phi^\prime_{8k+3}\rangle)=-\frac{1}{\sqrt{2}}(|1,0,0,...,0\rangle_a+|0,1,1,...,1\rangle_a)\otimes|0\rangle_{all}
\end{equation}
can be generated.
In order to test the effectiveness of our proposal,  we consider specifically, for example,
the case of five atoms. In Fig.~7, we plot the time-evolution behaviors of the occupation probability in the initial state $|\phi'_1\rangle$ governed by total Hamiltonian (Eq.~(\ref{total-n})) and by the effective Hamiltonian (Eq.~(\ref{eff-n})).
It is shown that the numerical results under these two Hamiltonians agree with each other reasonably well. Therefore, our effective model is valid.

\section{Discussions and conclusions}
The experimental feasibility of the proposed scheme is briefly analyzed as follows.
Practically, the atomic configuration involved in our scheme can be implemented with the $^{87}$Rb atom, whose relevant atomic levels are shown in Fig.~8.
\begin{figure}[b]
\begin{center}
\includegraphics[width=0.65\textwidth]{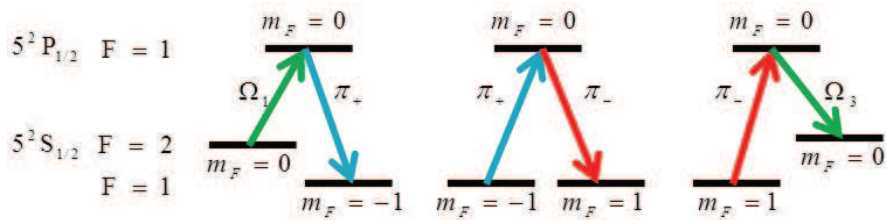}
\end{center}
\caption{Experimental atomic configuration used to generate tripartite GHZ state. Here, $\Omega_1$ and $\Omega_3$ are the classical fields, and $\pi_+$ and $\pi_-$ are the quantized cavity modes with different polarizations, respectively.}
\end{figure}
Where the first (third) atom is coupled resonantly to an external $\pi$-polarized classical field and a $\pi_+$ ($\pi_-$) polarized photon modes of the cavity, and the middle atom is coupled resonantly to
a $\pi_+$ and a $\pi_-$ polarized modes.
Based on the above discussions, in order to obtain fidelity larger than 90\%, one should keep the decay rate $\gamma<0.0055g$. This condition can be satisfied in recent experiments~\cite{37,38}, wherein for an optical cavity with the wavelength about $850$ nm the parameters are set as: $g/2\pi = 750$ MHz, $\gamma/2\pi = 2.62$ MHz, $\kappa_c/2\pi = 3.5$ MHz. Also, a near perfect fiber-cavity coupling with an efficiency larger than 99.9\% can be realized using fiber-taper coupling to high-Q silica microspheres~\cite{39}. The fiber loss at the $852$ nm wavelength is about $2.2$ dB/km \cite{40}, which corresponds to the fiber decay rate $1.52\times10^5$ Hz. With these parameters, it seems that the present entanglement-generation scheme with a high fidelity larger than 93\% could be feasible with the present experimental technique. The duration of the Zeno pulses $\Omega_1$ and $\Omega_3$ utilized to generate the desirable three-atom GHZ state can be estimated as $\pi/N_1\sqrt{\Omega_1^2+\Omega_3^2}\approx1.43\times10^{-8}$s, which is easily implemented for the present laser technology.

In conclusion, we have proposed an approach to achieve three-atom GHZ states by taking advantage of quantum Zeno dynamics. In present proposal, only one step operation is required to complete the generation.
The generation of the GHZ states is controllable, as the proposed quantum Zeno dynamics can be switched on/off
by controlling the classical drivings of the atoms at two ends.
We note also that, the generated GHZ state is no longer evolving, at least theoretically, after switching
off the classical drivings of the atoms at two-side cavities.
Additionally, the evolution of system is always kept in the subspace with null-excitation of cavity fields. So, the scheme is immune to the decay of cavity modes. Moreover, the scheme loosens the requirement of strong cavity-fiber coupling. We also show that the proposal can be easily generalized to prepared the N-atom entanglement without increasing the number of the applied classical fields.

\section*{Acknowledgments}
This work was supported in part by the National Natural
Science Foundation of China, under Grants No. 90921010
and No. 11174373, and the National Fundamental Research
Program of China, through Grant No. 2010CB923104.

%\begin{thebibliography}{000} %for 3 digits
%\begin{thebibliography}{00}  %for 2 digits


\begin{thebibliography}{0}    %for 1 digit

\bibitem{1} Introduction to Quantum Computation and Information, edited by H.-K. Lo, S. Popescu, and T. Spiller (World Scientific, Singapore, 1997).

\bibitem{2} M. A. Nielsen and I. L. Chuang, Quantum Computation and Quantum Information (Cambridge University Press, Cambridge, 2000).

\bibitem{3} C. H. Bennett and D. P. DiVincenzo,
%``Quantum information and computation''
Nature (London) {\bf 404}, 247 (2000).

\bibitem{4} A. K. Ekert,
%``Quantum cryptography based on Bell's theorem''
Phys. Rev. Lett. {\bf 67}, 661 (1991).

\bibitem{ghz} D. M. Greenberger, M. A. Horne, A. Shimony and A. Zeilinger,
%``Bell's theorem without inequalities''
Am. J. Phys {\bf 58}, 1131 (1990).

\bibitem{hbb} M. Hillery, V. Buzek, and A. Berthiaume,
%``Quantum secret sharing''
Phys. Rev. A {\bf 59}, 1829 (1999).

\bibitem{ghz_t1} X.-B. Zou, K. Pahlke, and W. Mathis,
%``Conditional generation of the Greenberger-Horne-Zeilinger state of four distant atoms via cavity decay''
Phys. Rev. A {\bf 68}, 024302 (2003).

\bibitem{ghz_t2} K. Pahlke, X.-B. Zou, and W. Mathis,
%``The generation of the Greenberger-Horne-Zeilinger state of four distant atoms conditioned on cavity decay ''
J. Opt. B Quantum Semiclass. Opt. {\bf 6}, S142 (2004).

\bibitem{ghz_t3} D. Gonta, S. Fritzsche, and T. Radtke,
%``Generation of four-partite Greenberger-Horne-Zeilinger and W states by using a high-finesse bimodal cavity''
Phys. Rev. A {\bf 77}, 062312 (2008).

\bibitem{ghz_t4} L. F. Wei, Y. X. Liu, and F. Nori,
%``Generation and Control of Greenberger-Horne-Zeilinger Entanglement in Superconducting Circuits''
Phys. Rev. Lett {\bf 96}, 246803 (2006).

\bibitem{ghz_e1} R. J. Nelson, D. G. Cory, and S. Lloyd,
%``Experimental demonstration of Greenberger-Horne-Zeilinger correlations using nuclear magnetic resonance''
Phys. Rev. A {\bf 61}, 022106 (2000).

\bibitem{ghz_e2} D. Leibfried et al.,
%``Creation of a six-atom `Schr\"{o}dinger cat' state''
Nature (London) {\bf 438}, 639 (2005).

\bibitem{ghz_e3} M. Neeley et al.,
%``Generation of three-qubit entangled states using superconducting phase qubits,''
Nature (London) {\bf 467}, 570 (2010).

\bibitem{qed} J.M. Raimond, M. Brune, and S. Haroche,
%``Manipulating quantum entanglement with atoms and photons in a cavity''
Rev. Mod. Phys. {\bf 73}, 565 (2001).

\bibitem{cz} J. I. Cirac and P. Zoller,
%``Preparation of macroscopic superpositions in many-atom systems,''
Phys. Rev. A {\bf 50}, R2799 (1994).

\bibitem{hl} J. Hong and H.-W. Lee,
%``Quasideterministic Generation of Entangled Atoms in a Cavity,''
Phys. Rev. Lett. {\bf 89}, 237901 (2002).

\bibitem{czkm} J. I. Cirac, P. Zoller, H. J. Kimble, and H. Mabuchi,
%``Quantum State Transfer and Entanglement Distribution among Distant Nodes in a Quantum Network''
Phys. Rev. Lett. {\bf 78}, 3221 (1997).

\bibitem{ecz} S. van Enk, J. Cirac, and P. Zoller,
%``Ideal Quantum Communication over Noisy Channels: A Quantum Optical Implementation,''
Phys. Rev. Lett. {\bf 78}, 4293 (1997).

\bibitem{bkpv} S. Bose, P. L. Knight, M. B. Plenio, and V. Vedral,
%``Proposal for Teleportation of an Atomic State via Cavity Decay,''
Phys. Rev. Lett. {\bf 83}, 5158 (1999).

\bibitem{lssh} S. Lloyd, M. S. Shahriar, J. H. Shapiro, and P. R. Hemmer,
%``Long Distance, Unconditional Teleportation of Atomic States via Complete Bell State Measurements,''
Phys. Rev. Lett. {\bf 87}, 167903 (2001).

\bibitem{pk} A. S. Parkins and H. J. Kimble,
%``Position-momentum Einstein-Podolsky-Rosen state of distantly separated trapped atoms,''
Phys. Rev. A {\bf 61}, 052104 (2000).

\bibitem{p} T. Pellizzari,
%``Quantum Networking with Optical Fibres''
Phys. Rev. Lett {\bf 79}, 5242 (1997).

\bibitem{19} A. Serafini, S. Mancini, S. Bose,
%``Distributed Quantum Computation via Optical Fibers''
Phys. Rev. Lett. {\bf 96}, 010503 (2006).

\bibitem{20} P. Peng, F.-L. Li,
%``Entangling two atoms in spatially separated cavities through both photon emission and absorption processes''
Phys. Rev. A {\bf 75}, 062320 (2007).

\bibitem{21} Z.-Q. Yin, F.-L. Li,
%``Multiatom and resonant interaction scheme for quantum state transfer and logical gates between two remote cavities via an optical fiber''
Phys. Rev. A {\bf 75}, 012324 (2007).

\bibitem{22} S.-Y. Ye, Z.-R. Zhong, S.-B. Zheng,
%``Deterministic generation of three-dimensional entanglement for two atoms separately trapped in two optical cavities''
Phys. Rev. A {\bf 77}, 014303 (2008).

\bibitem{23} Z.-B. Yang, S.-Y. Ye, A. Serafini, S.-B.Zheng,
%``Distributed coherent manipulation of qutrits by virtual excitation processes''
J. Phys. B: At. Mol. Opt. Phys. {\bf 43}, 085506 (2010).

\bibitem{24} X.-Y. Lv, L.-G. Si, X.-Y. Hao, and X. Yang,
%``Achieving multipartite entanglement of distant atoms through selective photon emission and absorption processes,''
Phys. Rev. A {\bf 79}, 052330 (2009).

\bibitem{25} S. B. Zheng,
%``Generation of Greenberger-Horne-Zeilinger states for multiple atoms trapped in separated cavities,''
Eur. Phys. J. D {\bf 54}, 719 (2009).

\bibitem{zl} A. Zheng, and J. Liu,
%``Generation of an N-qubit Greenberger-Horne-Zeilinger state with distant atoms in bimodal cavities''
J. Phys. B: At. Mol. Opt. Phys. {\bf 44}, 165501 (2011).

\bibitem{26} X.-Y. Lv, P.-J. Song, J.-B. Liu, and X. Yang,
%``N-qubit W state of spatially separated single molecule magnets''
Opt. Express {\bf 17}. 14298 (2009).

\bibitem{27} P.-B. Li, and F.-L. Li,
%``Deterministic generation of multiparticle entanglement in a coupled cavity-fiber system''
Opt. Express {\bf 19}, 1207-1216 (2011)

\bibitem{zeno1} P. Facchi, V. Gorini, G. Marmo, S. Pascazio and E. C. G. Sudarshan,
%``Quantum Zeno dynamics''
Phys. Lett. A {\bf 275}, 12 (2000).

\bibitem{zeno2} P. Facchi, S. Pascazio,
%``Quantum Zeno Subspaces''
Phys. Rev. Lett. {\bf 89}, 080401 (2002).

\bibitem{zeno3} P. Facchi, G. Marmo and S. Pascazio,
%``Quantum Zeno dynamics and quantum Zeno subspaces''
J. Phys: Conf. Ser. {\bf 196}, 012017 (2009).

\bibitem{zeno4} P. Facchi, S. Pascazio, A. Scardicchio and L. S. Schulman,
%``Zeno dynamics yields ordinary constraints''
Phys. Rev. A {\bf 65}, 012108 (2002).

\bibitem{state1} A. Luis,
%``Quantum-state preparation and control via the Zeno effect''
Phys. Rev. A {\bf 63}, 052112 (2001).

\bibitem{state2} X. B. Wang, J. Q. You and F. Nori,
%``Quantum entanglement via two-qubit quantum Zeno dynamics''
Phys. Rev. A {\bf 77}, 062339 (2008).

\bibitem{me} W.-A. Li, and G.-Y. Huang,
%``Deterministic generation of a three-dimensional entangled state via quantum Zeno dynamics''
Phys. Rev. A. {\bf 83}, 022322 (2011).

\bibitem{gate1} A. Beige, D. Braun, B. Tregenna and P. L. Knight,
%``Quantum Computing Using Dissipation to Remain in a Decoherence-Free Subspace''
Phys. Rev. Lett. {\bf 85}, 1762-1765 (2000).

\bibitem{gate2} J. D. Franson , B. C. Jacobs and T. B. Pittman,
%``Quantum computing using single photons and the Zeno effect''
Phys. Rev. A. {\bf 70}, 062302 (2004).

\bibitem{gate3} X.-Q. Shao, L. Chen, S Zhang and K.-H. Yeon,
%``Fast CNOT gate via quantum Zeno dynamics''
J. Phys. B: At. Mol. Opt. Phys. {\bf 42}, 165507 (2009).

\bibitem{gate4} J. D. Franson, T. B. Pittman and B. C. Jacobs,
%``Zeno logic gates using microcavities''
J. Opt. Soc. Am. B {\bf 24}, 209 (2007).

\bibitem{wei-prb-2005} L. F. Wei, Yu-xi Liu, and F. Nori, Phys. Rev. B {\bf 72}, 104516 (2005).

\bibitem{37} S. M. Spillane, T. J. Kippenberg, K. J. Vahala, K. W. Goh, E. Wilcut, H. J. Kimble,
%``Ultrahigh-Q toroidal microresonators for cavity quantum electrodynamics''
Phys. Rev. A {\bf 71} 013817 (2005).

\bibitem{38} J. R. Buck, H. J. Kimble,
%``Optimal sizes of dielectric microspheres for cavity QED with strong coupling''
Phys. Rev. A {\bf 67} 033806 (2003).

\bibitem{39} S. M. Spillane, T. J. Kippenberg, O. J. Painter, K. J. Vahala,
%``Ideality in a Fiber-Taper-Coupled Microresonator System for Application to Cavity Quantum Electrodynamics''
Phys. Rev. Lett. {\bf 91} 043902 (2003).

\bibitem{40} K. J. Gordon, V. Fernandez, P. D. Townsend, G. S. Buller,
%``A short wavelength GigaHertz clocked fiber-optic quantum key distribution system''
IEEE J. Quantum Electron. {\bf 40} 900 (2004).





\end{thebibliography}
\end{document}